\documentclass[aps, twocolumn, superscriptaddress,
longbibliography]{revtex4}

\usepackage{graphicx}
\usepackage{amsmath}
\usepackage{amssymb}
\usepackage{graphicx}
\usepackage{afterpage}
\usepackage{amsfonts}
\usepackage{amssymb}
\usepackage{graphicx}
\usepackage{grffile}
\usepackage[T1]{fontenc}
\usepackage[utf8x]{inputenc}

\usepackage{braket}
\usepackage{textcomp}
\usepackage{xcolor}
\usepackage{float}
\usepackage{tabularx}
\usepackage{booktabs}
\usepackage{soul}
\usepackage{verbatim} %for block comments

\begin{document}
\title{Photoluminescence efficiency of MBE-grown MoSe$_2$ monolayers featuring narrow excitonic lines and diverse grain structures.}

\author{Mateusz~Raczy\'nski}
\email{mateusz.raczynski@fuw.edu.pl}\affiliation{Faculty of Physics,
University of Warsaw, ul. Pasteura 5, 02-093 Warszawa, Poland}

\author{Julia~Kucharek}
\affiliation{Faculty of Physics, University of Warsaw, ul. Pasteura
5, 02-093 Warszawa, Poland}

\author{Kacper~Oreszczuk}
\affiliation{Faculty of Physics, University of Warsaw, ul. Pasteura
5, 02-093 Warszawa, Poland}

\author{Aleksander~Rodek}
\affiliation{Faculty of Physics, University of Warsaw, ul. Pasteura
5, 02-093 Warszawa, Poland}

\author{Tomasz~Kazimierczuk}
\affiliation{Faculty of Physics, University of Warsaw, ul. Pasteura
5, 02-093 Warszawa, Poland}

\author{Rafa\l{}~Bo\.zek}
\affiliation{Faculty of Physics, University of Warsaw, ul. Pasteura
5, 02-093 Warszawa, Poland}

\author{Takashi~Taniguchi}
\affiliation{Research Center for Materials Nanoarchitectonics, National Institute for Materials Science,  1-1 Namiki, Tsukuba 305-0044, Japan}

\author{Kenji~Watanabe}
\affiliation{Research Center for Electronic and Optical Materials, National Institute for Materials Science, 1-1 Namiki, Tsukuba 305-0044, Japan}

\author{Wojciech~Pacuski}
\affiliation{Faculty of Physics, University of Warsaw, ul. Pasteura
5, 02-093 Warszawa, Poland}

\author{Piotr~Kossacki}
\affiliation{Faculty of Physics, University of Warsaw, ul. Pasteura
5, 02-093 Warszawa, Poland}

%%%%%%%%%%%%%%%%%%%%%%%%%%%%%%%%%%%%%%%%%%%%%%%%%%%%%%%%%%%%%%
\begin{abstract}

Recent studies have demonstrated that using h-BN as a substrate for the growth of transition metal dichalcogenides can significantly reduce excitonic linewidths. However, many other optical parameters still require optimization. In this work, we present a detailed study of the low-temperature photoluminescence efficiency of MBE-grown MoSe$_2$ monolayers on h-BN substrates, comparing them to state-of-the-art exfoliated monolayers encapsulated in h-BN. We demonstrate that a quantitative comparison between samples requires accounting for interference effects and Purcell enhancement or suppression of the emission. By accounting for these effects in both photoluminescence and Raman signals, we show that the overall intrinsic luminescence efficiency is proportional to the sample coverage. Consequently, we find that exciton diffusion and edge effects are negligible in spectroscopy of MBE-grown samples, even for nanometer-sized crystals.

\end{abstract}
%%%%%%%%%%%%%%%%%%%%%%%%%%%%%%%%%%%%%%%%%%%%%%%%%%%%%%%%%%%%%%

\date{\today}

\maketitle
%%%%%%%%%%%%%%%%%%%%%%%%%%%%%%%%%%%%%%%%%%%%%%%%%%%%%%%%%%%%%%

\section{Introduction}

The scalable growth of large-area epitaxial materials is of key importance for any applications involving semiconductor two-dimensional crystals such as transition metal dichalcogenides (TMDs). Among various growth techniques, chemical vapour deposition (CVD) is frequently reported and typically results in a good flatness and structural quality of the achievable crystals due to the high growth temperatures regarding WS$_2$ \cite{Okada_ACSNano_2014, Carozo_SA_2017, Orsi_N_2018}, MoS$_2$ \cite{Wang_ACSNano_2015, Yan_NL_2015, Chae_APL_2017, Zhang_PRB_2019, Uchiyama_NPG2D_2019, Shree_2D_2019} or WSe$_2$ \cite{Zhang_NL_2018, Zhang_ACS_2019, Delhomme_APL_2019}. Although molecular beam epitaxy (MBE) of TMDs  (WSe$_2$ \cite{Jeong_PRB_2018, Liu_2D_2015, Nakano_NL_2017}, MoS$_2$ \cite{Fu_JACS_2017}, MoTe$_2$ \cite{Vishwanath_JMR_2016, Roy_ACSAMI_2016, Vishwanath_JCG_2018, Ogorzalek2020, Seredynski2022} or MoSe$_2$ \cite{Shimada_JSAP_1990,Ohuchi_JAP_1990,Xenogiannopoulou_Nanoscale_2015,Liu_2D_2015,Zhang_NC_2016,Vishwanath_JMR_2016,Roy_ACSAMI_2016,Onomitsu_APE_2016,Chen_ASCNano_2017,Dau_APL_2017,Wei_Nano_2017,Poh_adma2017,Poh2018,Jeong_PRB_2018,Dau_APL_2019}) is still in an early stage of development, its relatively low growth temperatures offer potential for application on diverse substrate structures, including those containing fragile semiconductor devices and ultra-flat surfaces with low sticking coefficients. Recently, large, homogeneous structures with excellent optical properties have been successfully grown using MBE \cite{Pacuski2020,Ludwiczak2021,Ludwiczak2024,Vergnaud2024} thanks to use of hexagonal boron nitride (h--BN) substrates. Despite these advancements, epitaxial techniques face challenges related to the low mobility of transition metals on the substrate surface \cite{Mortelmans2020,Seredynski2022}, which first leads to the formation of small grains that later merge into continuous layers \cite{Fu_JACS_2017,Pacuski2020}. Consequently, epitaxial TMD layers suffer from dislocations resulting from grain merging\cite{Zhang_NL_2018,Calderon2022} and high amount of point defects \cite{Poh2018}. Both grain boundaries and point defects strongly influence the photoluminescence spectrum, particularly by broadening the excitonic emission lines \cite{Vergnaud_2D_2019,Pacuski2020,Kucharek2023} and significantly reducing the lifetime of the excitonic states by enabling non--radiative relaxation channels \cite{Polczynska2023,Oreszczuk2024}.

The narrowest exciton lines of MBE-TMDs are obtained for MoSe$_2$ monolayers grown on mechanically exfoliated h-BN \cite{Pacuski2020, Vergnaud2024}. A single wafer used in such growth typically contains several flakes of varying thickness on a Si/SiO$_2$ substrate. Alternatively, using Metal-Organic Chemical Vapor Deposition (MOCVD) it is possible to produce wafer-scale h--BN with homogenous thicknesses which acts as an almost equally efficient substrate for MBE growth of MoSe$_2$ \cite{Ludwiczak2021,Ludwiczak2024}. In both cases, the resulting TMD structures exhibit high spatial homogeneity in the shape of the photoluminescence spectra, both within individual h-BN flakes and across macroscopic distances on the scale of centimeters for a given sample \cite{Pacuski2020}. Test of this homogeneity is limited by the optical spatial resolution, as photoluminescence can only resolve features (inhomogeneities) larger than the diffraction limit, determined by the excitation beam's wavelength and the numerical aperture of the setup. Other techniques, such as cathodoluminescence \cite{doi:10.1021/acs.nanolett.7b03585} performed on monolayers or scanning near--field optical microscopy \cite{KimKim_2021_3397_3415}, come as sample-invasive or very challenging experimentally. Consequently, most of the optical measurements involve signal--averaging over a micrometer-sized spot, while the sizes of the MBE--grown MoSe$_2$ grains typically remain below the diffraction limit, in the range of tens of nanometers.

This work focuses on the mechanisms responsible for reducing photoluminescence intensity, particularly exploring the impact of interference effects, Purcell enhancement, small crystal grains, and homogeneously distributed defects. We analyze low--temperature excitonic spectra within a series of MBE--grown MoSe$_2$ samples with varying grain sizes. Moreover, we explain how to fairly compare the photoluminescence (PL) and Raman scattering intensities from different samples to properly account for the effects of interferences due to the photonic environment resulting from underlying substrate \textit{i.e.} SiO$_2$ and h--BN layer of varying thicknesses and very thick (opaque, 450\hspace{2pt}$\mu$m) silicon wafer. We show that the intensity of photoluminescence is, first of all, dependent on excitation laser interferences and luminescence interferences itself (Purcell effect). By accounting for this photonic environment and monolayer coverage for all measured spots, it was possible to assess intrinsic optical quality (brightness) for all investigated samples. Effectively, we could compare the photoluminescence quantum yield of MBE--grown samples against the sample that was exfoliated and embedded between layers of the encapsulating h--BN.

\begin{figure}[h]
\begin{center}
\includegraphics[width=0.97\columnwidth]{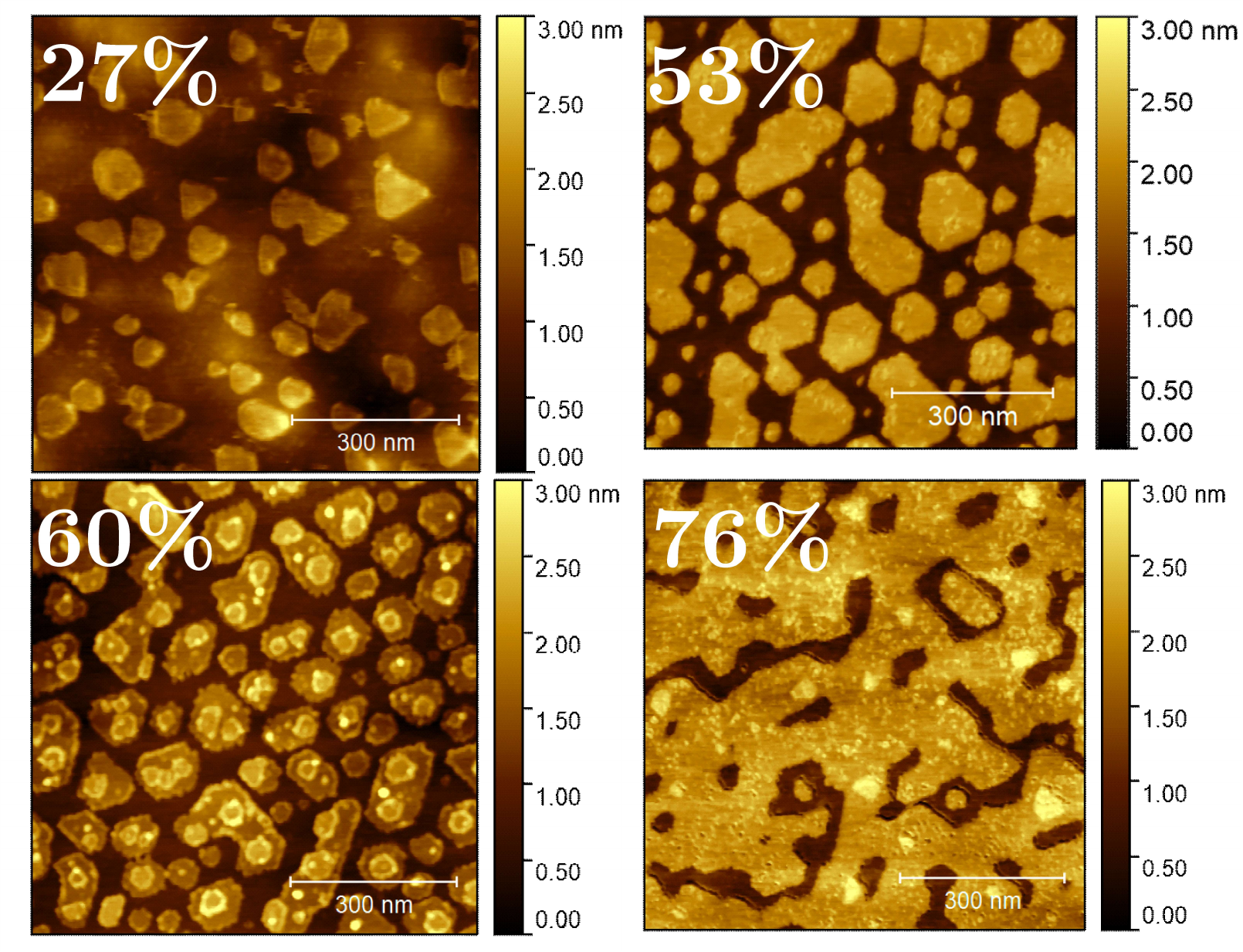}
\end{center}
\caption[]{The picture shows AFM height maps of four samples grown in separate MBE processes. The maps provide the coverage factor of a monolayer, and information regarding morphology, flakes orientation, size or overgrowth with the bilayer on top of the monolayer. The sides of the square maps are all 800\hspace{2pt}nm in dimension.}
\label{fig:F1_Sample_geometry_and_morphology_800nm_4samples}
\end{figure}

\section{\textbf{Samples}}

The samples studied in this work were fabricated using molecular beam epitaxy (MBE). They were grown on h-BN substrates prepared using flakes exfoliated from bulk h-BN onto a commercially available oxidized silicon wafer with a 90\,nm silicon dioxide thickness. These h-BN flakes -- in micro--scale -- serve as high-quality, flat and smooth substrates without dangling bonds and uncompensated electric charge, enabling the best growth of molybdenum diselenide monolayers.
 
The process in ultra--high vacuum ($10^{-9}$ Torr) started with a 15--minute degas of the wafer at 800$^{\circ}$C and subsequent cooling to about 300$^{\circ}$C to initiate growth by opening both the Se and Mo shutters.
Growth was performed with a relatively small selenium flux, with beam equivalent pressure (BEP) of about $10^{-7}$ Torr, and an even smaller Mo flux leading to the growth of incomplete MoSe$_\textrm{2}$ monolayer during 15 hours of deposition. Selenium with 99.99999\% purity (7N) was deposited from a standard effusion cell while Mo was deposited from an e-beam source with a Mo rod with 99.995\% purity (4N5). After growth, samples were annealed at 800$^{\circ}$C in higher Se flux (BEP about $10^{-6}$ Torr), with Mo shutter closed. The precise amount of the deposited material was determined after growth by analysis of images obtained with atomic force microscopy (AFM), as presented in Fig.~1. Further, we use these coverage percentage values to refer to the samples. In particular, for the sample presented in Fig.~1a there is 27\% monolayer coverage. This sample was grown with 10\,h of MoSe$_2$ deposition and 2\,h of the post--growth annealing. For the next 3 samples shown in Figs.~1b,c,d monolayer coverage is respectively 53\%, 60\%, and 76\%. These 3 samples were grown with 15\,h of MoSe$_2$ deposition and 3 annealings, each 2\,h long: two intermediate (after 5\,h and after 10\,h) and one at the end of the process. The difference between processes originates, therefore, from the fluctuation of molecular fluxes and resulting differences in the deposition and the re--evaporation of material. As we have discussed in \cite{Pacuski2020}, very slow growth and multiple annealing are helpful for the growth of quasi-continue monolayers, as shown in Fig.~1d. However, different strategies can be used to optimize optical properties, as we discuss at the end of the manuscript.

The grain structure of epitaxial MoSe$_2$ is well resolved on AFM images (Fig.~1). The characteristic sizes of the MoSe$_\textrm{2}$ monolayer grains are on the order of tens of nanometers. The scans clearly show that the growth follows a substrate-predefined orientation, as most of the sub-micron MoSe$_\textrm{2}$ flakes are aligned in the same direction, exhibiting the six-fold rotational symmetry characteristic of TMDs. However, grains are too small to be resolved in optical experiments due to the diffraction limit of a laser spot - about 500\,nm. In the micrometre scale, MBE MoSe$_2$ is almost perfectly homogenous, as discussed in Refs \cite{Pacuski2020,Ludwiczak2021,Polczynska2023,Ludwiczak2024}. In this work, we use large--scale homogeneity to probe the effect of h--BN thickness on optical properties. The thickness of the h--BN flakes varied due to the inherent randomness of the exfoliation process. Therefore, it provided a crucial benefit of statistical assembly with a wide spread of h--BN thicknesses for the same (almost identical) MoSe$_2$ monolayer. It would be not possible with means of mechanical exfoliation to produce at once so many similar MoSe$_2$ layers lying on different h--BN flakes. The thicknesses of studied h--BN flakes ranged from just a few nanometers to nearly 500\hspace{2pt}nm, as determined by AFM.

A single sample (wafer) is a result of a single growth process, therefore the only nominal difference across different spots within the same sample is the photonic environment, specifically the thickness of the h--BN flake at a given spot.  In contrast, different samples were grown in separate MBE processes with varied parameters and are expected to differ. Indeed, AFM maps show differences in the shape, morphology, and coverage of the MoSe$_\textrm{2}$ monolayers, which are presented in Fig.~1. 

While MoSe$_2$ grew on both the amorphous silica and h--BN substrates, only the monolayers on the boron nitride exhibited satisfactory quality worth further investigation.

For comparison, we also studied an exfoliated sample. This sample consisted of an exfoliated MoSe$_\textrm{2}$ monolayer encapsulated with h-BN flakes on both the bottom (92\hspace{2pt}nm) and top (18\hspace{2pt}nm), deposited on the same type of Si/SiO$_2$ substrate that was used for the MBE-grown samples. This particular exfoliated sample, produced using the dry PDMS stamp method, has already been the subject of optical studies \cite{RodekNanoPho21} and is known to be of high quality at specific points on this flake.

\section{Experimental techniques}

The optical study presented here includes low--temperature photoluminescence and room temperature anti--Stokes Raman scattering. The low--temperature PL measurement has been done in helium--flow cryostat model \textit{Janis ST--500} in $T=10\,$K. The excitation source was the \textit{COBOLT Samba 100\,mW $\lambda=532\,$nm} CW semiconductor laser. The excitation power measured just before the microscope objective was set to match approximately $P=200\,\mu$W. The micrometre resolution necessary to probe the MoSe$_\textrm{2}$ monolayer grown on a particular h--BN flake was achieved with the use of the Nikon microscope L Plan achromatic objective with x100 magnification, $\textrm{NA}=0.7$, and working distance $\textrm{WD}=6.5\,$mm. The spectrometer used was \textit{Andor SR-500i} with diffraction grating 600/500nm and Peltier cooled CCD camera of \textit{Andor DV420-FI}. The Raman spectroscopy measurements were performed using a commercial \textit{Renishaw inVia Raman Microscope} equipped with $532\,$nm laser.

The thicknesses of h--BN flakes -- at which positions the optical measurements of MoSe$_2$ were done -- have been determined by the AFM height map. The AFM model is \textit{Bruker Dimension Icon} and it was used in the PeakForce Tapping mode. This directly determined height was then cross--checked against the reflectivity spectra (Fig.~4) -- measured in low--temperature conditions at the same time with the PL. The exact h--BN thicknesses enabled us to correlate the influence of the interference effects on the measured intensity of the PL and the Raman spectra. That influence is explained in the next chapter. As mentioned above, the AFM scans at large magnification provided us with the monolayer coverage factor on each investigated sample Fig.~1.

\begin{figure}[h]
\begin{center}
\hspace*{-15pt}\includegraphics[width=1.1\columnwidth]{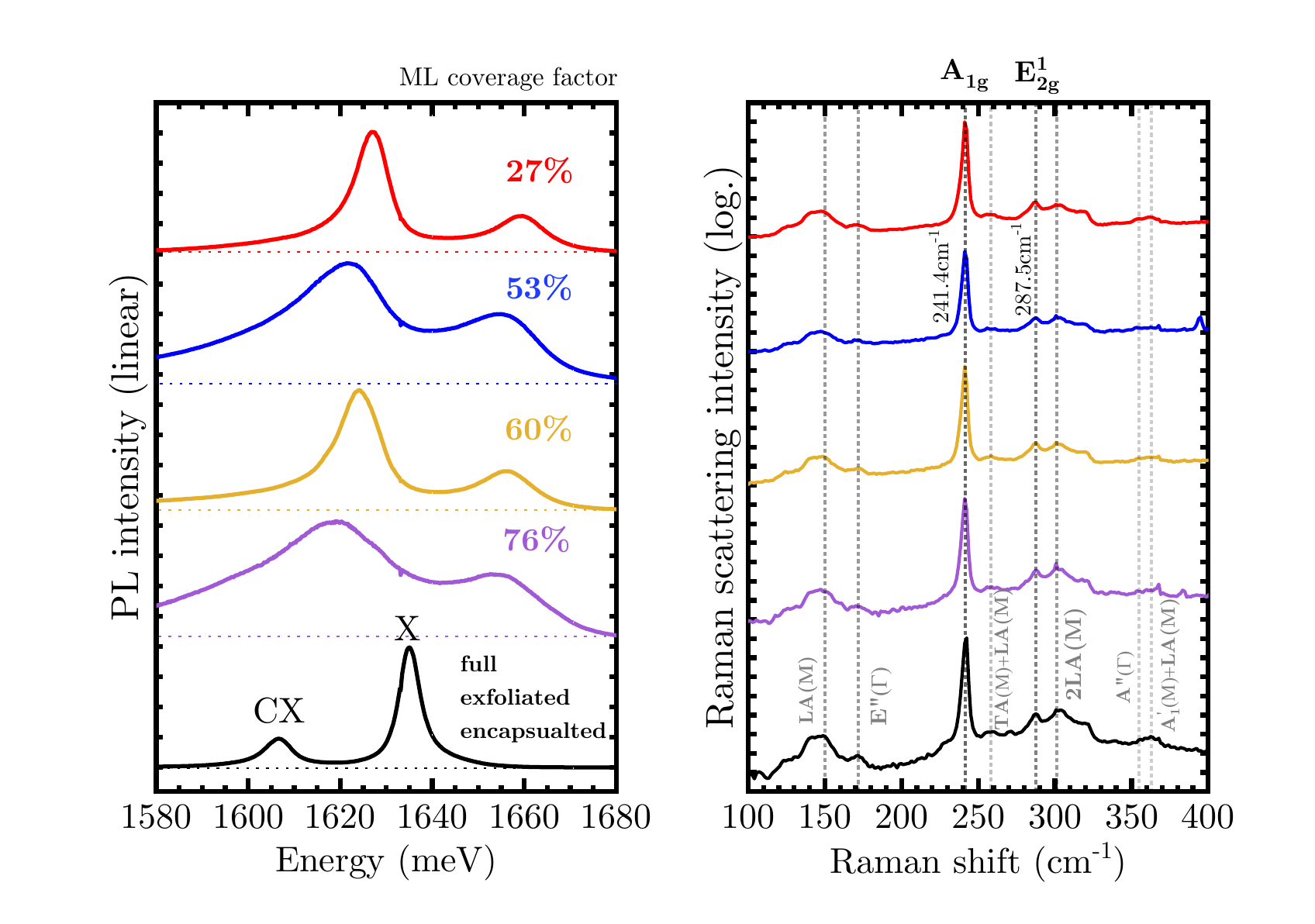}
\end{center}
\caption[]{Representative spectra of the photoluminescence (T=10\hspace{2pt}K) and the Raman scattering (at room temperature) of all investigated MBE--samples and the reference exfoliated MoSe$_{\textrm{2}}$ monolayer encapsulated from the top and the bottom with h--BN flakes. The spectra were normalised to similar maximum intensity for easier shape comparison. In both cases, the laser used for PL excitation and Raman scattering was green $\lambda = 532$\hspace{2pt}nm laser.}
\label{fig:F2_Representative_spectra_offset_PL_and_Raman_selected_ordered}
\end{figure}

\section{Optical spectra}

Fig.~2 presents PL and Raman scattering spectra measured for four MoSe$_2$ samples with different ML coverage. The PL normalised spectra consisting of two distinct peaks are presented on the left--hand panel of the figure. The lower energy peak is the signature of the charged exciton (CX) and the higher energy peak corresponds to neutral A--exciton (X$^\textrm{A}$ or simply X). The ratio of oscillator strengths of the charged exciton to the neutral one suggests intrinsic doping of MBE--grown samples. In particular, this ratio increases with the ageing of the sample due to air exposure. Such exposure to the air for a few hours was the time spent on measuring the Raman spectra. The low--temperature PL spectra before and after this exposure to ambient conditions showed a slow but systematic shift to a higher CX/X intensities ratio across every sample across their whole areas.
In contrast, the exfoliated and encapsulated sample persisted in a mostly neutral state of a significantly weaker charged exciton peak in the overall PL spectrum. We attribute the red--shift of the exfoliated sample spectrum to the MBE--grown samples to additional strain and screening of the dielectric environment induced by the top h--BN layer. This distinction has also been observed in the energy difference between neutral and charged exciton yielding approximately $31\,$meV for MBE--grown sample and $28\,$meV for the exfoliated one.

On the right-hand side of the panel, we also present the normalized Raman spectra of the samples. The peaks corresponding to the particular vibrational modes \cite{Tonndorf:13} have been marked on the graph. For further analysis, we focused on the intensity of the most pronounced $A_{1g}$ mode at approximately $241.4\,\textrm{cm}^{-1}$. The FWHM of this Raman line remained on average $2.5\,\textrm{cm}^{-1}$ for each sample.

\begin{figure}[h]
\begin{center}
\hspace*{-15pt}\includegraphics[width=1.1\columnwidth]{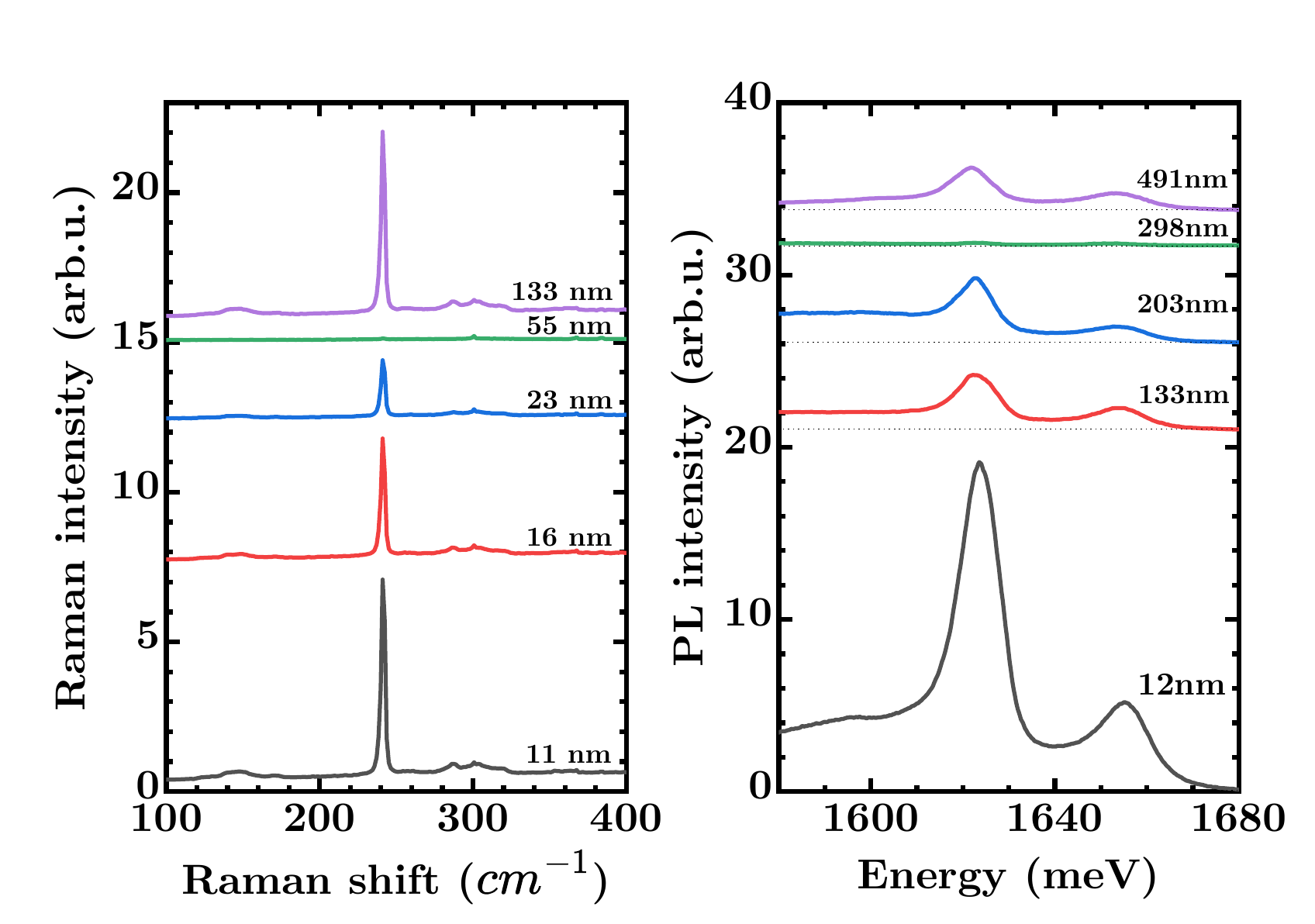}
\end{center}
\caption[]{ Raman scattering and PL spectra measured on the same sample with 60\% ML coverage. The spectra  were measured from flakes with varying h--BN thickness, as indicated near each spectrum. The interference effects significantly influence the spectral emission intensity but not its shape. The scale is linear but for clarity, the spectra were shifted verically. The integrated data from all flakes for both panels is presented in Fig.~5.}
\label{fig:F2b_representative_Raman_and_PL_spectra_vs_thickness}
\end{figure}

\section{Light intensity comparison}

Direct comparison of the PL intensity of different samples does not explicitly reflect the intrinsic properties of samples, as quantitative analysis of light emission intensity from complex structures requires accounting for interference effects and Purcell enhancement or suppression of the emission.\cite{Oreszczuk2024, transfermatrix, Clua-Provost2024NanoLetters} To develop and test a model incorporating these factors, we took advantage of the inherent random distribution of the thicknesses of h--BN flakes obtained through mechanical exfoliation. In fact, the photoluminescence (PL) intensity within the same sample, measured on several h--BN flakes of varying thickness, exhibited even greater variability than the one observed between different samples on h--BN spots of similar colour and therefore similar h--BN thickness (Fig.~3). To assess these effects, we analyzed the PL and Raman scattering intensities across multiple flakes for every sample grown.

\section{Interferences effects resulting from h--BN thickness}

To tackle the issue of interferences, we performed modelling using the transfer--matrix formalism \cite{transfermatrix}. The light intensity was averaged for S and P light polarization and over all the angles within the acceptance angle cone defined by the numerical aperture of the used microscope objective. The refractive indexes of the different materials utilized in the calculations were as follows: SiO$_2$ (fused silica)\cite{Malitson_SiO2_1965}; Si \cite{Schinke_Si_2015}; h--BN (taking only n(o))\cite{Grudinin_BN_2023}. These data have not been corrected for the temperature change from room to cryogenic temperature. However, the relevant change in the reflective index would be too small to affect the interferences significantly.

While the dielectric function of monolayer MoSe$_2$ has been already established in the literature \cite{dielectricmose2}, in order to account for possible growth-specific variations, we opted to rely on data collected for the same sample series. Specifically, the measured reflectivity spectrum was used as the reference for fitting the modelled reflectivity spectrum by adjusting the monolayer dielectric function. The dielectric response of the MoSe$_2$ monolayer was modelled by a series of Lorentz oscillators for excitonic transitions and additional higher energy frequency--independent dielectric constant. The neutral $X^\textrm{A}$ ($\sim$1658\,meV -- 27\% ML sample) and $X^\textrm{B}$ ($\sim$1865\,meV -- 27\% ML sample) excitons were visible in the reflectance spectrum of each sample. The CX exciton was observed in the PL spectrum but not in the reflectance.
We also did not include the band--band absorption in the model. The issue of \textit{porosity} or -- more precisely --  the \textit{coverage factor} of the MoSe$_2$ monolayer was simulated by interpolating between the refractive index of MoSe$_2$ (full monolayer) and the one of the air (vacuum) $n=1$. The sizes of MoSe$_2$ flakes are smaller than the optical resolution of the setup at both excitation and emission light wavelengths. Hence, the averaging approach is justified.

\begin{figure}[h]
\begin{center}
\includegraphics[width=0.7\columnwidth]{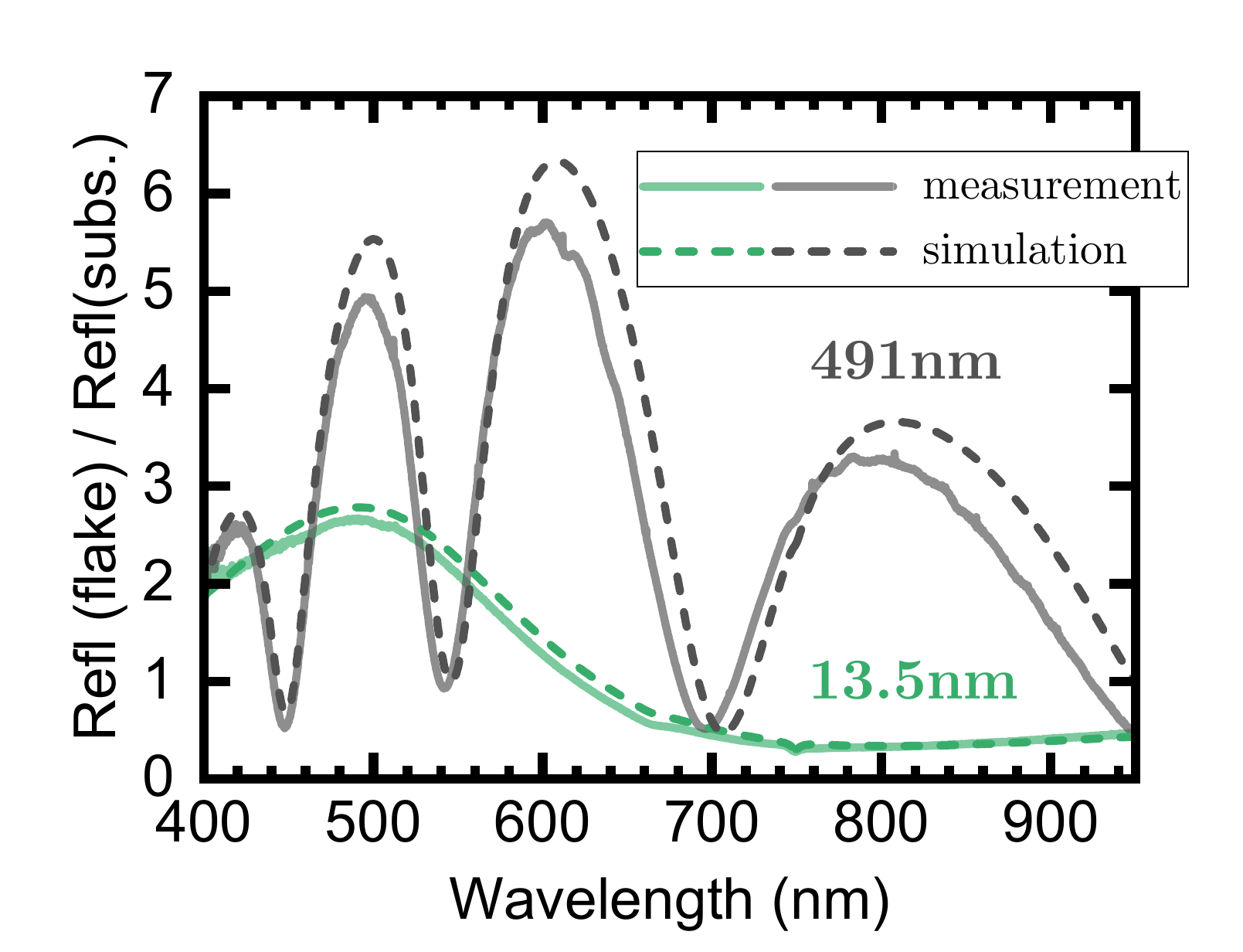}
\end{center}
\caption[]{The graph presents measured and calculated by transfer--matrix formalism reflectivity from the h--BN flakes with thickness 13.5\,nm and 491\,nm, and with MoSe$_2$ on its top. The reflectivity spectra are normalised by the reflectivity measured from just a simple substrate of Si and SiO$_2$ of the same oxide thickness. The lower intensity of the experimental data might be caused by an additional scattering factor due to the unevenness of h--BN flakes.}
\label{fig:F2_reflectivty_spectra_representative}
\end{figure}

\begin{figure}[h]
\begin{center}
\includegraphics[width=1.00\columnwidth]{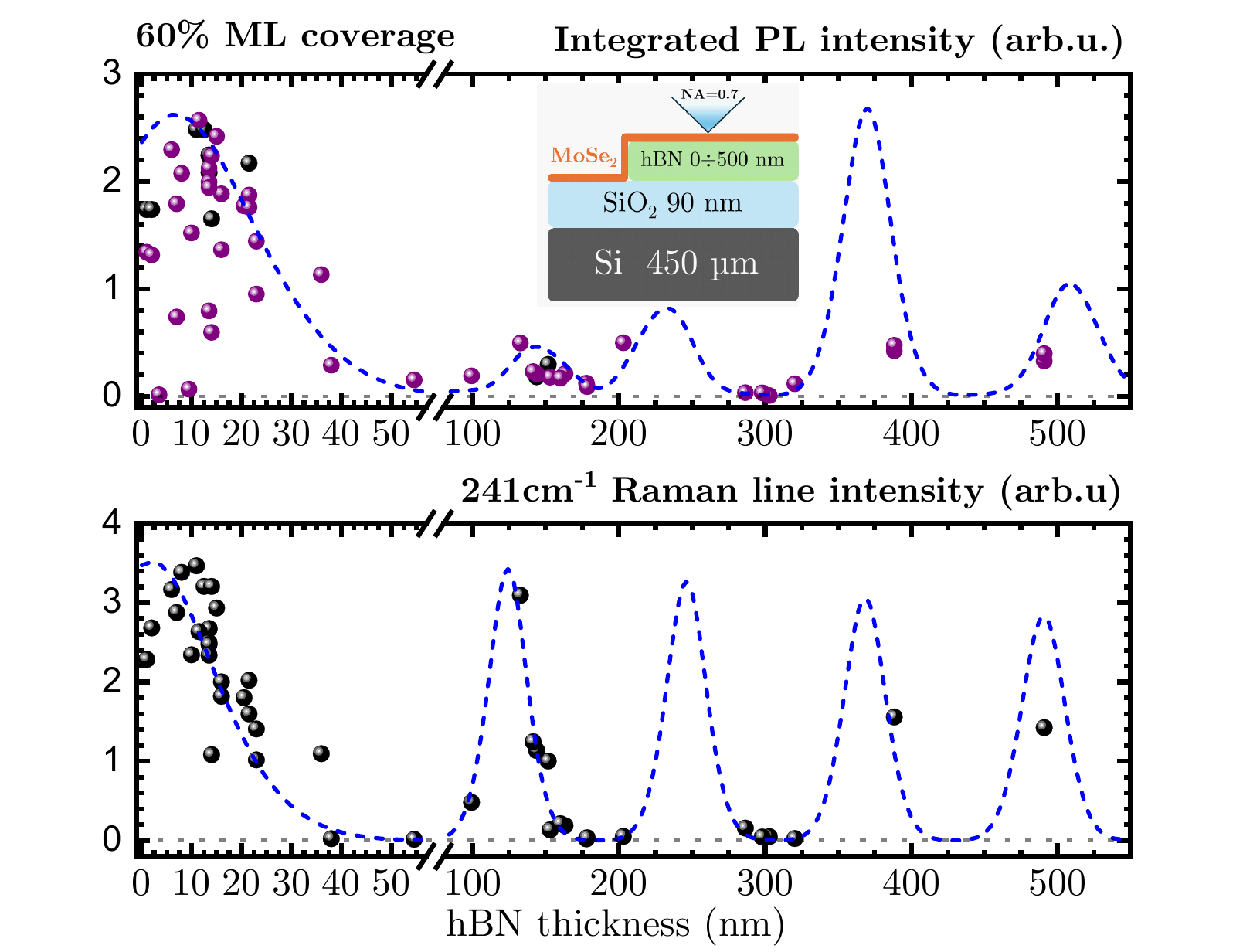}
\end{center}
\caption[]{The upper panel (a) shows the data points of measured PL intensity (integrated spectrum) vs the thickness of the h--BN flake at which position (spot) the measurement was taken. The dashed line depicts the theoretical calculation obtained by the transfer matrix method averaged by the objective numerical aperture. The lower panel (b) analogically presents the data point for the Raman scattering for line 241\,cm$^\textrm{-1}$.}
\label{fig:F3_interferences_effects_in_PL_and_Raman_Scattering_and_structure}
\end{figure}

Within such a computational approach, we calculated the laser intensity at the emitter position -- MoSe$_\textrm{2}$ monolayer -- to find the dependence of the laser interferences on the h--BN thickness. The laser intensity experienced by the emitter directly impacts the resulting intensity of the PL or the Raman signal. 

Similarly, the analogous dependence was calculated for the emitted PL light and the scattered one during the Raman measurement. In the latter case, the intensity of the signal measured in a given geometry is simply directly proportional to the interference factor, similarly, as in the case of the incoming laser intensity \cite{raman_thin_films}. However, in the case of the PL signal the effect of the light interferences is more subtle, depending on the presence of competing non--radiative recombination channels. It is well known and shown for layered materials, that the thicknesses influence the Purcell factor and thereby change the emission lifetimes \cite{FangPRL19}. Upon favourable interference conditions, \textit{i.e.}, when the outgoing photon mode significantly contributes to the local density of states at the MoSe$_2$ monolayer position, the spontaneous emission rate is enhanced. In the absence of other radiative and non--radiative recombination processes, the enhanced emission rate would not change the observed PL intensity, as the total number of photons would equal the total number of the created excitons. However, when non--radiative channels open, they start to compete with the radiative ones, so the intensity integrated over time will decrease by a factor $\tau_\gamma^{-1}/(\tau_\gamma^{-1}+\tau_{nr}^{-1})$, where $\tau_\gamma^{-1}$ is the rate of the spontaneous emission and the $\tau_{nr}^{-1}$ is the rate of all other non--radiative relaxation channels. The larger the Purcell factor is, the faster the spontaneous emission becomes and more energy is radiated out of the structure during the lifetime of the excitons. In the case of the TMDs, especially the ones grown by epitaxy, the majority of relaxation pathways occur via non--radiative channels with sub-picosecond time--scales as it has been shown by Oreszczuk \textit{et al.} \cite{Oreszczuk2024}. The radiative lifetimes measured with a streak camera in best samples can reach a few picoseconds for neutral exciton and usually dozens of picoseconds for the charged one \cite{FangPRL19} with the record--breaking result close to the 300\,ps \cite{doi:10.1021/acsnano.1c04331}. Therefore, in our case, in the above--mentioned fraction we can assume that the overall exciton lifetime $\tau=(\tau_\gamma^{-1}+\tau_{nr}^{-1})^{-1}$ is dominated by non-radiative processes and therefore the intensity factor is simply directly proportional to the Purcell factor $\textrm{F}_\textrm{P}$, which in turn stems from the interferences in the structure similarly to earlier discussed case of incoming laser or the outgoing Raman photons.

The combined effects of interferences are presented together with the experimental data in the Fig.~5. The emitted light was assumed to be exactly $750\textrm{\hspace{2pt}nm}$ and $525\textrm{\hspace{2pt}nm}$ accordingly. This translates to approx. CX and X emission peaks and $\textrm{A}_\textrm{1g}$ Raman line with energy shift of $241\,\textrm{cm}^\textrm{-1}$ -- in anti--Stokes scattering.
The model curve nicely explains the observed differences in the intensity of the signal at different spots, proving the importance of the interference effects, and further confirming the dominating contribution of non--radiative relaxation regime in case of the PL.
In the following section, we will therefore account for different photonic environments by dividing the raw measured signal by the predicted enhancement factor, which will expose the intrinsic material differences between different samples.

\section{The PL and Raman emission efficiency}

Fig.~2 contains a collection of the PL (a) and Raman (b) spectra of the four investigated samples with different monolayer coverage measured under the same experimental conditions. The relative intensity measured within each sample (Fig.~3) varies mostly due to interference effects (Fig.~5), which is problematic in terms of feedback for further growth optimization.

Yet, as we discussed in the previous section, we account for these differences by dividing the optical signal measured at a particular spot on a given sample by the theoretical enhancement factor calculated as outlined above. We note that the enhancement factor is calculated separately for each spot, based on h--BN thickness determined by AFM measurements for this particular spot. The more time--saving approach would be to determine the h--BN thickness from reflectivity spectra, but it lacks the spatial resolution and the certainty characteristic for the AFM.

After accounting for the influence of interference effects on the emitted light intensity for each of the MBE processes, the PL brightness of each wafer was correlated with the ML coverage factor obtained from the AFM maps, which ranged from 27\% to 76\%. These results are shown in Fig.~6a. The corrected brightness of the MBE--grown samples was expressed in relation to the similarly corrected signal from the reference exfoliated MoSe$_2$. It is important to note that the exfoliated sample was encapsulated in h--BN from both the top and bottom, while the MBE--grown samples were not covered from the top, creating a systematic comparative difference, particularly in the energy of the excitonic lines in the PL (Fig.~2).

\begin{figure}[h]
\hspace*{-10pt}\includegraphics[width=1.1\columnwidth]{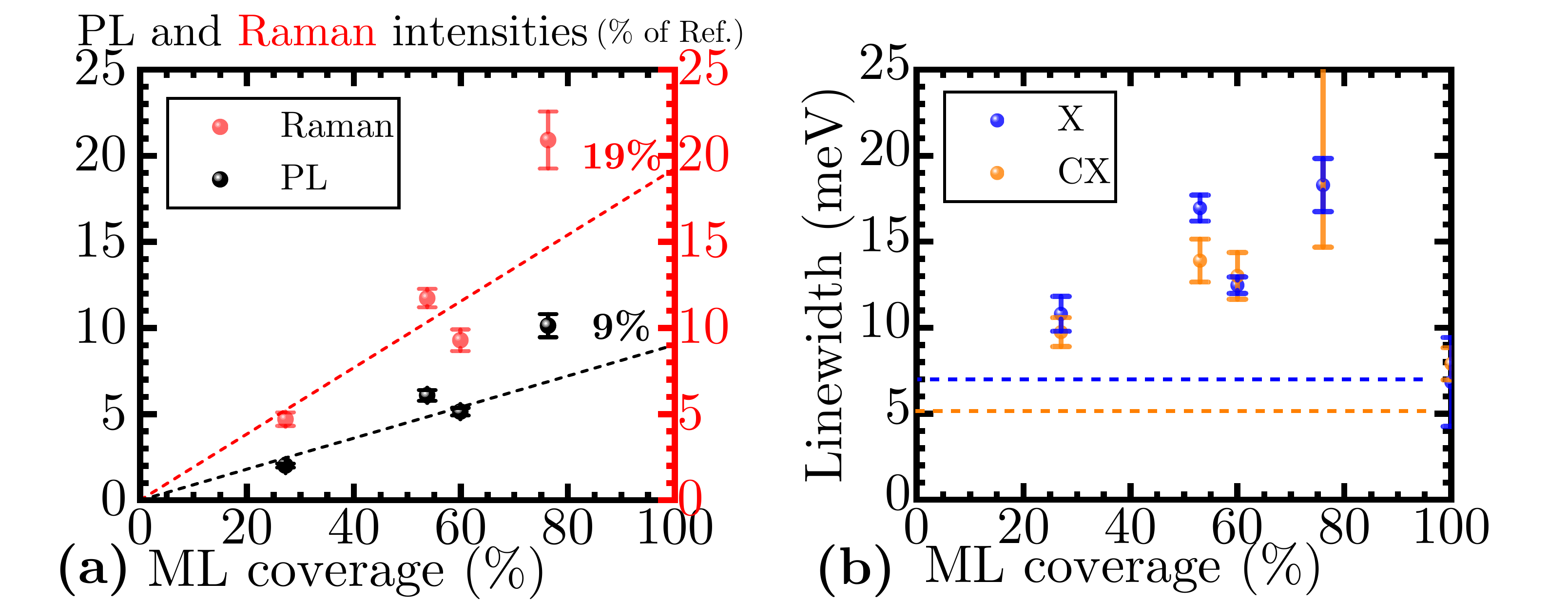}
\caption[]{The plot depicts the first generation of samples, their intrinsic total PL and Raman $A_{1g}$ line intensities after accounting for interference effects vs the monolayer coverage factor calculated with AFM height maps (Fig.~1) \\ The larger size of flake islands leads to broadening the excitonic lines visible on the PL spectrum. Even the presence of a bilayer has a lower impact on the features visible on the spectrum than the sizes of MoSe$_2$ flakes. It is especially striking when one compares the samples with 53\% and 60\% ML coverage factors. The horizontal dashed lines depict the narrowest CX (red) and X (black) peaks measured for the exfoliated and encapsulated reference MoSe$_2$ monolayer.}
\label{fig:F4_average_PL_and_Raman_intensities_normalised_1st_series}
\end{figure}

The data shows that the intrinsic PL brightness changes systematically within the studied series of samples, exhibiting a nearly linear trend versus the coverage factor, as seen in Figure 6a. Extrapolation of this trend to 100\% of ML coverage yields a value of approximately 9\% of the exfoliated sample PL intensity, which translates to an 11-fold faster non-radiative decay rate in the case of the MBE samples.
A similar linear trend is observed for the Raman signal. Here, the Raman scattering intensity extrapolated to 100\% ML coverage reaches 19\% of the Raman signal for the exfoliated sample, which is a notably higher ratio than the one for the PL signal.

In terms of the intrinsic efficiency of the optical signal, the MBE--grown monolayers are therefore still inferior in comparison with the exfoliated samples. The better performance of the Raman signal than of the PL signal can be explained in the following manner. Raman scattering is sensitive to the very local crystal environment, while Wannier--Mott excitons are delocalized over distances spanning several crystal unit cells. Before their phonon-mediated relaxation to the light cone occurs, they may travel longer distances either ballistically or diffusively, probing larger crystal volumes. If an exciton encounters a crystal defect, it can localize or relax via non-radiative processes. Consequently, excitons are much more sensitive to crystal disorder -- each probes a larger volume. High non-radiative relaxation rates may also occur due to the proximity of flake edges, to which excitons can potentially diffuse. Given the small nanometer--sized flakes, the near-edge regions could -- in principle -- contribute significantly to the overall surface, and, if the diffusion lengths are comparable to the flake sizes, non-radiative processes would be substantially enhanced.

From this perspective, comparing samples with larger and smaller crystal grain sizes, such as those corresponding to 53\% and 60\% monolayer coverage, is particularly insightful. Both cases exhibit quite similar brightness despite the different grain sizes (Fig.~2). Although a sample with 60\% coverage with smaller grains exhibits a lower brightness, it could also stem from the significant bilayer overgrowth and not necessarily due to the edge proximity.
Thus, we conclude that diffusion and edge effects are of lesser importance. This statement is even more pronounced by Fig.~7 which shows the linear proportionality of the PL and Raman intensities. Unlike the PL from excitonic states, the Raman scattering signal is independent of diffusion effects. Hence, if the proportionality occurs for samples of different grain sizes and the ML coverage then the diffusion of excitonic complexes is negligible. A small saturation of this effect can be observed for the sample with the most robust optical response with 37\% of ML coverage. We will discuss the case of this sample next.

\begin{figure}[h]
\begin{center}
\includegraphics[width=0.7\columnwidth]{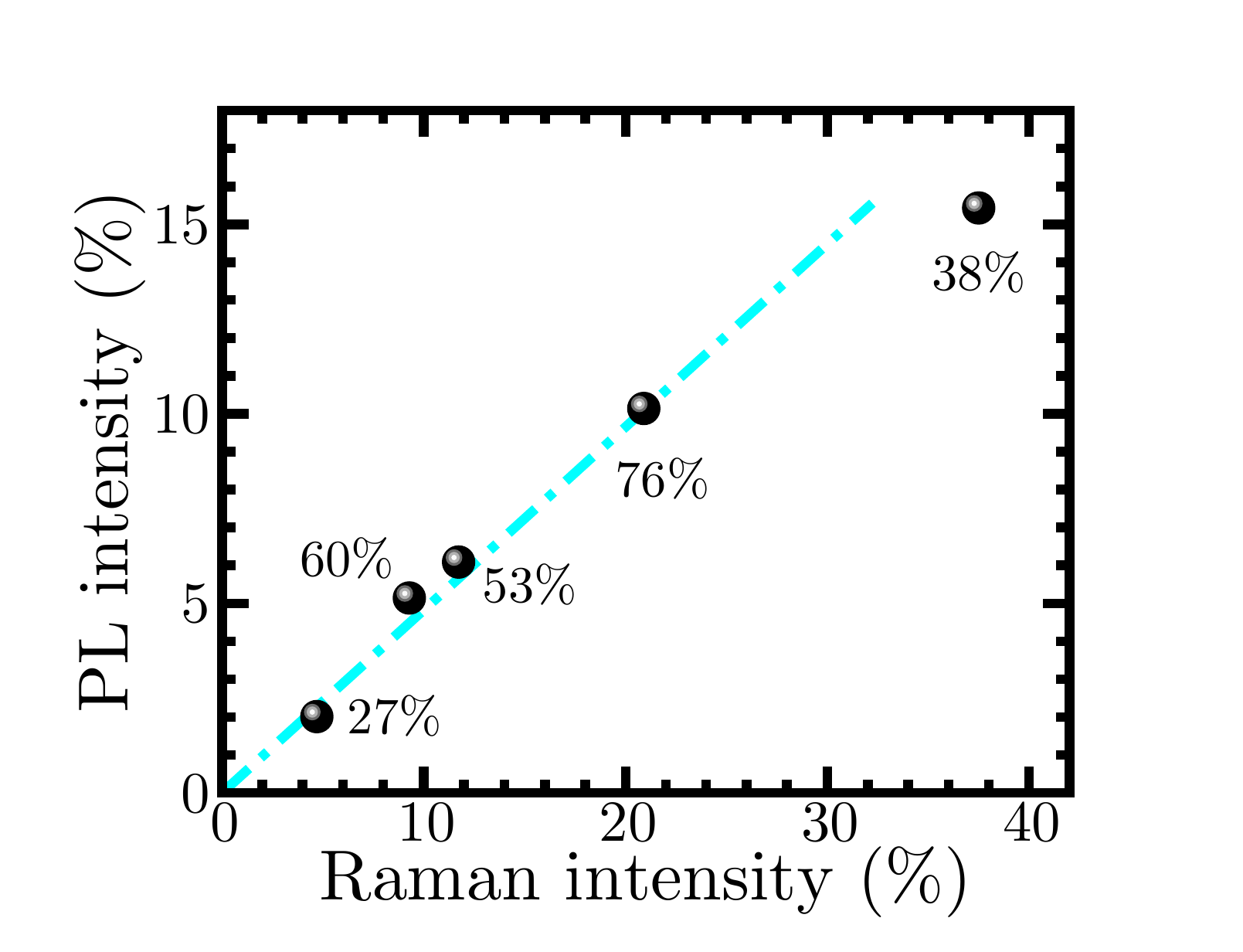}
\end{center}
\caption[]{ The PL and Raman intrinsic brightnesses (as a percentage of the reference exfoliated sample) of 5 different samples. All points are annotated with their respective ML coverage. }
\label{fig:F8_PL_vs_Raman_percent}
\end{figure}

However, as grain size and coverage increase, a notable broadening of the linewidth is observed (Fig.~6b). We interpret this as resulting from strain induced by the difference in thermal expansion coefficients of MoSe$_2$ flakes and the substrate. Such strain is easily relaxed in small, isolated grains but becomes significant when the grains merge into bigger flakes and finally into a continuous layer.

\begin{figure}[h]
\begin{center}
\includegraphics[width=0.7\columnwidth]{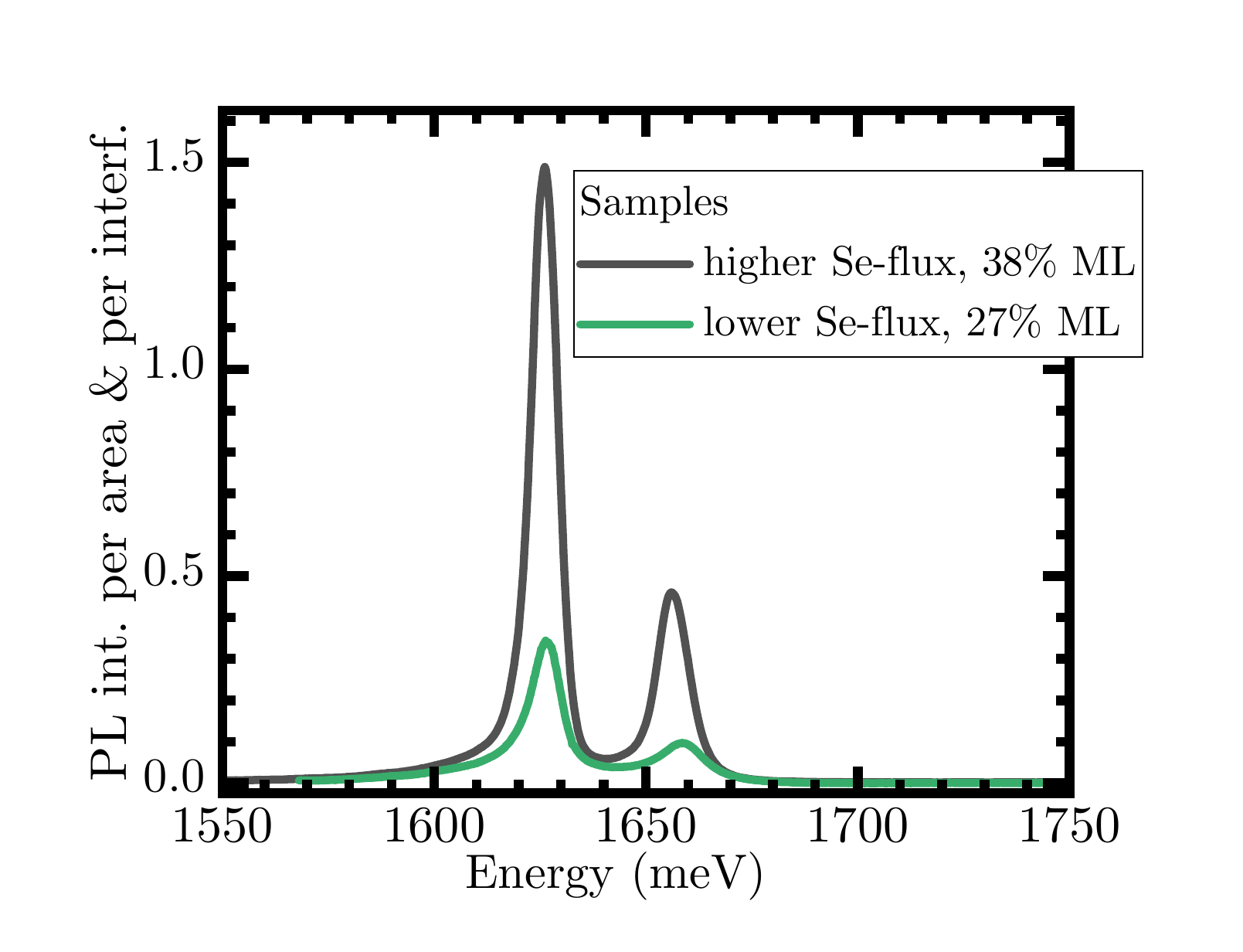}
\end{center}
\caption[]{ The graph presents two PL spectra (arb.u.) normalised by the ML coverage and the interference effect from the photonic environment. Therefore, the difference in intensity of both spectra should be attributed to the samples' intrinsic ratio of radiative to non--radiative relaxation rates due to the different growth conditions. The green curve depicts PL from the previous generation of samples. The black line is the new one grown with excess selenium flux. The integrated intensity is approximately 5--fold between these spectra. }
\label{fig:F7_normalised_PL_spectra_new_vs_old_generation_samples}
\end{figure}

\begin{figure}[h]
\begin{center}
\includegraphics[width=0.55\columnwidth]{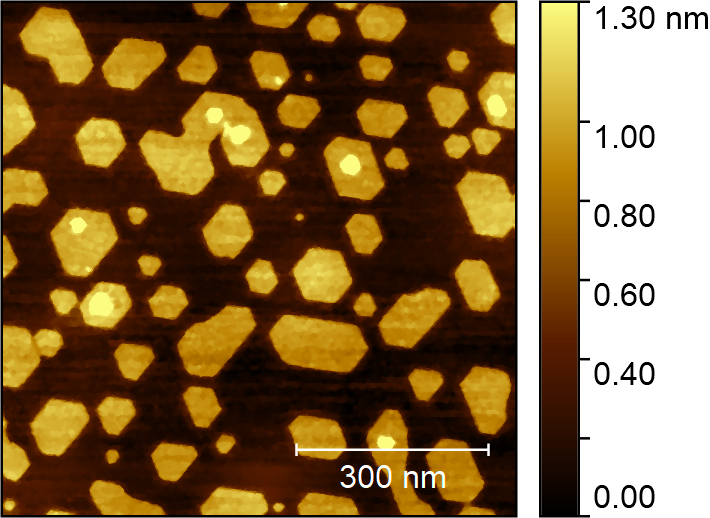}
\end{center}
\caption[]{ The AFM image of the sample grown with high selenium flux and 38\% ML coverage. All the edges exhibit fine alignment with six main directions, even the bilayer overgrowth still follows this behaviour confirming a perfect epitaxial growth.}
\label{fig:F7_UW2156_02_0_8um}
\end{figure}

Finally, we note that although the studied series of samples presents a uniform behaviour, as proven by linear dependence in Figs.~6a and 6b, the deduced intrinsic efficiency should not be treated as the absolute limit attainable for the MBE-grown monolayers. On the contrary, identifying two main side factors -- namely the interferences due to a particular h--BN thickness and only partial monolayer coverage -- gave us a tool to compare the efficiency of different samples reliably. This data can be used as feedback to steer the further growth optimization. As a proof-of-concept, we compare there another sample, which was grown in a separate process under growth conditions optimized for enhancement of PL intensity. The main difference, when comparing it to previously analysed samples was the selenium flux which was one order of magnitude larger for both growth and annealing. Due to Se consumption, such high Se flux cannot be maintained for longer than a few hours. Therefore Mo source has been also recalibrated and consequently, 80--minute deposition time was enough to achieve monolayer coverage of 38\%. Such a sample yields a significantly stronger PL signal, as illustrated in Fig. \ref{fig:F7_normalised_PL_spectra_new_vs_old_generation_samples}, reaching 44\% of the PL yield of the exfoliated sample, after accounting for the incomplete monolayer coverage and the individual interference factor. The Raman signal yield is as intense as for the exfoliated sample -- within the limits of the measurement and the normalisation accuracy. Spectral lines are a little narrower (CX -- 7.2\,meV and X$^\textrm{A}$ -- 7.6\,meV) than in the sample with the lowest ML coverage. Also, the AFM imaging shows better alignment between different grains (Fig. \ref{fig:F7_UW2156_02_0_8um}). The improvement of the intrinsic PL brightness due to the increase of Se flux during growth and annealing indicates that the main factor responsible for non--radiative recombination is related to Se vacancies or additional Mo atoms attached to MoSe$_2$. Overall, the improved performance of the new sample is direct proof of how important is to obtain reliable feedback information about the intrinsic PL yield, without obscuring factors like optical interferences and varying monolayer coverage.

\section{Summary}
We have presented optical studies of MoSe$_2$ monolayer samples grown by Molecular Beam Epitaxy, confirming that they exhibit macroscale spatial homogeneity in the shape of photoluminescence spectra but significantly differ in intensity at different spots. We identify that this variation, observed also in Raman scattering within a given epitaxial process, stems mostly from the interference effects caused by variations in the thickness of the underlying h--BN substrate. These effects include laser interference and the Purcell effect, where the exciton radiative lifetime is shortened due to emitter--light mode coupling in a regime dominated by non--radiative decay channels. After accounting for this factor, we find that the intrinsic brightness of different samples is proportional to the monolayer coverage rather than the MoSe$_2$ grain size, leading us to conclude that diffusion and edge effects are negligible in MBE--grown samples.

We argue that accounting for the identified main factors, i.e., the optical interferences and the coverage factor, is the crucial element for further growth optimization of such structures.

\section*{Acknowledgements}
This work is supported by  National Science Centre, Poland, under projects 2020/39/B/ST3/03251, 2021/41/B/ST3/04183, and 2023/49/N/ST11/04125. K.W. and T.T. acknowledge support from the JSPS KAKENHI (Grant Numbers 21H05233 and 23H02052), the CREST (JPMJCR24A5), JST and World Premier International Research Center Initiative (WPI), MEXT, Japan.

\bibliography{PPAR}

\end{document}